\documentclass[12pt]{article}
\usepackage{fullpage}

\begin{document}
\input{amssym.def}

\newcommand{\mod}{{\rm mod}}
\newcommand{\br}{{\Bbb R}}
\newcommand{\bz}{{\Bbb Z}}
\newcommand{\bt}{{\Bbb T}}
\newcommand{\bc}{{\Bbb C}}
\newcommand{\lr}{\longrightarrow}
\newcommand{\ch}{{\cal H}}
\newcommand{\co}{{\cal O}}
\newcommand{\ct}{{\Gamma}}
\newcommand{\cv}{{\cal V}}
\newcommand{\ol}{\overline}
\newcommand{\wh}{\widehat}
\newcommand{\lan}{\langle}
\newcommand{\ran}{\rangle}

\title{\Large \bf The Poincar\'e Group of Discrete Minkowskian
Space-Time}
\author{~~~~ \\ [.18in]
P.P. Divakaran \\
{\small Chennai Mathematical Institute,
92 G.N. Chetty Road, T. Nagar,
Chennai-600 017, India.}\\ 
and \\
{\small Institute of Mathematical Sciences,
C.I.T. Campus, Taramani,  Chennai-600 113, 
India} \\ [2mm]
{\small E-mail: ppd@cmi.ac.in}
}
\date{}
\maketitle


\noindent
{\bf Abstract}

The lattice of integral points of 4-dimensional Minkowski space,
together with the inherited indefinite distance function, is considered
as a model for discrete space-time.  The Lorentz and Poincar\'e groups 
of this discrete space-time are identified as subgroups of the
corresponding Lie groups.  The lattice Lorentz group has irreducible 
projective (including linear) representations which are restrictions
of (all) finite-dimensional irreducible projective representations of the
Lorentz Lie group and hence can be used to describe all integral and 
half-odd-integral helicity.  The (4-torus) momentum space has a 
well-defined ``light cone'' of null points and there are orbits of  
the lattice Lorentz group  
lying entirely in the torus light cone and having the lattice 
euclidean group of the plane as little group.    Wigner's
method for the Poincar\'e Lie group can then be adapted to show, in the
first instance, that the lattice Poincar\'e group has unitary 
representations describing lattice free fields of zero mass
and an arbitrary Lorentz helicity, in particular chiral fermions.
There are no representations with a nonzero invariant mass.

\newpage
\noindent
{\bf 1. Introduction}

The modelling  of space and space  - time by discrete  sets or lattices
has a long history, originating in efforts to regularise quantum field
theory [1-3].  Over the past  few decades, lattice methods have become
effective  tools in  several areas  of study:  abelian  and nonabelian
gauge models [4,5], quantum gravity  and strings [6,7], etc. etc. They
are  also an  indispensable part  of the  machinery  of nonperturbative
numerical  gauge theory  calculations [8,9].   In almost  all  of this
activity, the underlying ``continuum'' is euclidean space $(\br^3$) or
space-time $(\br^4)$ given a  euclidean metric, or manifolds which are
locally euclidean.  Not infrequently,  the lattices considered are not
regular.  These  features are  clearly inimical to  the implementation
and  exploitation  of the  natural  symmetries  of  realistic  space-time:  
random lattices  have  no  symmetries and  lattice
rotation groups in  euclidean space, in any number  of dimensions, are
finite groups.  This is an obvious handicap.  First, the restoration of
symmetries  in the continuum  limit becomes  a nontrivial  task.  More
seriously,  the  denial of  natural  symmetries  can  lead to  various
pathologies on the lattice.  (We shall have occasion to refer to some
troubles of this sort in the concluding section of this article).

The natural symmetries  that are the concern of  the present paper are
those  of flat minkowskian  space-time $\br^4$.   We describe  here the
first  results of a  study of  lattices embedded  in $\br^4$  with the
inherited  minkowskian   distance,  focusing  on   the  corresponding
(discrete)   Poincar\'e  group.  
Minkowskian  lattices have  seldom  been considered  in the  extensive
literature of  lattice physics and  never, to the  author's knowledge,
from  the  viewpoint  of  their symmetries.   Though  the  mathematical
methods required in this endeavour are less widely known than the 
theory of representations of Lie  groups and Lie algebras, there does  
exist a sufficiently rich and deep
body  of knowledge  on discrete  subgroups of  Lie groups  to  make 
the effort worthwhile.

To  see what  we  should be  aiming at,  we  have only  to recall  the
fundamental significance of relativistic  symmetries in the context of
quantum theory.  Not only do  they govern processes through the operation
of   conservation  laws   and  selection   rules,  but,   through  the
identification of  certain irreducible unitary  representations of the
Poincar\'e group  with 1-particle  states, they actually  {\it define}
elementary  particles; masses and helicities are invariants of 
representations and   the  functions   defining  the  space   of  a
representation are the corresponding (momentum space) free fields  [10-13].  
This is
the ideal  against which any proposal for  discretising space-time ought
to  be,  ultimately, judged.   Such  a  project  will consist  of  the
following  steps,  at the  least:  i)  identify  and characterise  the
discrete counterpart of the Poincar\'e group or, rather, its universal
covering  group;  ii) find  a  general  method  for constructing  its
irreducible unitary  representations; iii) show that some  at least of
these representations have  a satisfactory physical interpretation and
iv)  derive the  free-field dynamics  (field equations  and subsidiary
conditions).  The  first three of these questions  are addressed in this
paper and the  answers are,  in the context  of lattice kinematics,  more or
less satisfactory, with some surprises.  The lattice Poincar\'e group
is  easy  to determine  and  (the  ``2-fold  cover'' of)  its  Lorentz
subgroup turns out to be a  discrete group of an especially nice type. 
In  particular,  the  latter  has finite  dimensional  representations
which, for purposes of describing helicities, are as good as those of
the  Lorentz  (Lie) group.  
The momentum space  properties of the group, however, are
quite unlike  those in the continuum  case.  We shall find, apart from the 
inevitable momentum cutoff,  that it is
not  possible to  associate a  nonzero mass  to a  representation  in a
sensible way  - the  analogues of the  massive representations  of the
Poincar\'e (Lie) group have the  serious drawback that at certain values
of  the momenta,  all  of  them behave  like  tachyonic representations.  
Remarkably, massless representations (understanding the meaning of mass
and masslessness is  part of the work) are free  from this difficulty. 
(All statements  regarding momenta are  of course to be  understood in
the general context of lattice momenta which are defined and conserved
modulo  the inverse lattice  spacing).  
The construction of the most general massless representation is, as is 
true in general for discrete groups, much more demanding a task than in
the continuum case.  This effort is only initiated here, for certain
special types of ``momentum shells'' or orbits.
What  is notable is one consequence of  the combination 
of a good description of
(arbitrary)  Lorentz  helicity  and   of  masslessness:  chiral
fields are   automatic  on   the  Minkowskian  lattice,   in  fact
obligatory.  The  most serious drawback  of euclidean lattice  physics is
thus obviated by going minkowskian.

There are  of course crucial  differences in the physics  of continuum
and discrete  quantum relativity.  These  will be touched upon  in the
concluding section.   But it is possible to  argue, nevertheless, that
they  are not fatal  to the  tantalising possibility of the 
lattice model  of special relativity  being taken
seriously as describing physics at a high enough energy scale, 
{\it without actually going to the continuum limit}.

\noindent
{\bf 2. The Discrete Poincar\'e Group}

The group of isometries  of 4-dimensional Minkowski space $\br_M^4$ is
$O(3,1,\br) \vec{\times}  \br^4$, where $O(3,1,\br )$ is  the full (or
extended)  Lorentz  group,  $\br  ^4$  is the  translation  group  and
$\vec{\times}$  indicates the semidirect  product, the  arrow pointing
from the quotient subgroup to  the normal subgroup.  Of this, only the
subgroup connected to the identity  appears to be an exact symmetry of
nature.   We  shall denote  the  connected  Lorentz group  $SO(3,1,\br
)_{\rm conn}$ $=  SO(3,1,\br )/\{ \pm 1 \in  SO(S,1,\br )\}$ by $L(\br
)$,  the corresponding inhomogeneous  group $L( \br  ) \vec{\times}
\br^4$ by  $P(\br )$  and refer  to them simply  as the  (continuum or
real) Lorentz and Poincar\'e group.

The discrete  space-time of this  work is the hypercubic  lattice $\bz_M^4$
of points in  $\br^4$ with integer  coordinates (the
lattice  spacing is  the  unit of  length  and time),  with a  distance
function given by the metric  in $\br_M^4$: the (length)$^2$ of $X \in
\bz_M^4 =  \{ X_{\mu}  \in \bz  \mid \mu =0,\cdots  ,3\}$ is  $X_0^2 -
X_1^2 - X_2^2 - X_3^2 =  X_{\mu} X_{\mu} = X^2$.  The lattice Lorentz
group is then  the subgroup of $L(\br )$  obtained by restricting every
$4 \times 4$  matrix $\lambda \in L (\br )$  to have integral entries:
$L(\bz ) = SO( 3,1,\bz ) / \{ \pm 1 \in SO(3,1,\bz )\}$, where, as the
notation makes  clear, $SO(3,1,\bz  )$ is $SO(3,1)$  over the  ring of
integers.  $L(\bz )$  acts on  the discrete  translation  group $\bz^4$
exactly  as in  the corresponding  continuum case  and  the semidirect
product $P(\bz ) = L(\bz ) \vec{\times} \bz^4$, the discrete Poincar\'e
group,  is our  relativity group.   It is  an interesting  remark that
while $L(\bz  )$ is the  trivial group in  $1+1$ dimensions, it  is an
infinite group in all higher dimensional Minkowski lattices.

The representations of  $P(\bz )$ that are of  interest in the context
of  quantum  theory are  its  projective  unitary representations,  in
accordance  with  Wigner's  general  theorem on  symmetries.   In  the
continuum,  such  representations of  $P(\br  )$  are  found by  first
establishing   [10]   that    every   continuous   projective   unitary
representation   of  $P(\br   )$   lifts  to   a  continuous   unitary
representation of its universal covering group $\wh{P} (\br ) = \wh{L}
(\br ) \vec{\times} \br^4 = SL (2,\bc ) \vec{\times} \br^4$, $SL(2,\bc
)$ being the  universal cover of $SO(3,1,\br )_{\rm  conn}$.  This key
result,  which  is  clearly  very  specific to  Lie  groups,  has  the
following ingredients [10]: i) though $\br^4$ has nontrivial projective
representations,  they  do  not  extend  to $P(\br  )$  as  nontrivial
projective  representations; ii) though  semidirect product  groups $G
\vec{\times}  A$ with $A$ abelian can  have  projective representations  whose
restrictions  to $G$ and $A$ are  linear  representations ({\it  via}
1-cocycles on $G$ with values in the character group of $A$) this does
not happen  for $P(\br  )$ because $L(\br  )$ is semisimple;  and
iii) every  projective representation of  $L(\br )$ lifts to  a linear
representation of its universal cover, again because of semisimplicity.
To deal with  $P(\bz )$ with the same degree  of completeness will take
us  into  difficult  terrain   and,  for  our  purposes,  it  is
unnecessary.  Our  first aim being the  understanding of helicities,
which are  described by the  representations of the Lorentz  group, we
ignore ingredients i) and ii) (the validity of point i) for $P(\bz )$
is actually easy to establish) and concentrate on ingredient iii).  The
assertion iii) is a special case  of a general result which says that,
given  any group  $G$, we  can construct  a group  $\wh{G}$ determined
fully by $G$ with the  property that every projective representation of
$G$ lifts to a linear  representation of $\wh{G}$ [14].  $\wh{G}$ is called
a universal central extension of $G$ and is not always unique [14-16].
However,  a connected  semisimple  Lie group  has  a unique  universal
central extension  and it coincides  with its universal cover,  so that
equivalence classes of its  projective representations are classified by
the character group of its  fundamental group.  This is the reason why
$\wh{L} (\br )$ is $SL (2,\bc )$.  (Many Lie groups commonly occurring
in  physics  have  the   property  that  their  nontrivial  projective
representations have  no relationship whatever  with their fundamental
groups [15,16]).  A linear  representation of $\wh{L} (\br )$ restricting
to its centre $\bz_2$ ($=$ the  fundamental group of $L(\br )$) as the
trivial (nontrivial) character passes  to the quotient group $L(\br )$
as   a  trivial  projective, i.e. linear,   (nontrivial  projective)
representation  and   these  are  the  only   classes  of  projective
representations of $L(\br )$.  Keeping all this in mind, we shall refer to
$\wh{L} (\br )$ and $\wh{P} (\br )$ themselves as the Lorentz and Poincar\'e 
groups.

The  corresponding problem for  $L(\bz )$  is not  so neatly  settled. 
Indeed, lacking  a physically significant criterion of  continuity as is
available  for representations of  Lie groups,  the quest  for ``all''
projective representations of $L(\bz )$ is unduly ambitious, perhaps
ill-defined.  We shall  content ourselves with  showing the
existence  of  certain   finite  dimensional  (nonunitary)  projective
representations of  $L(\bz )$ (and, eventually,  of projective unitary
representations of  $P(\bz )$) which are inherited  naturally from those
of  $L(\br )$,  by  restriction -- in  general,  a nontrivial  projective
representation  of  a group  need  not restrict  to  a  subgroup as  a
nontrivial projective representation.

To  implement this  aim, we  look for  a subgroup  $\wh{L}(\bz  )$ of
$SL(2,\bc  ) =  \wh{L}(\br  )$  which is  such  that every  projective
representation  of $L(\bz  )$ that  is a  restriction of  a projective
representation  of  $L(\br )$  lifts  to  a  linear representation  of
$\wh{L} (\bz )$;  in other words, $\wh{L} (\bz )$  should be such that
its  centre contains $\bz_2$,  $\wh{L}(\bz )  / \bz_2  = L(\bz  )$ and
$L(\bz )$ is not  a subgroup of $\wh{L} (\bz )$.  This  can be done by
following the standard treatment (as given,  
for example, in [17]) of $L(\br )$
and $\wh{L}  (\br )$).  Denote  by $H(2,\bc )$  the (real) vector space  of $2
\times  2$   complex  matrices   which  are  hermitian and by $\tau_i$
($i=1,2,3$) the Pauli spin  matrices.  The association of $x \in \br_M^4$
to $x_{\mu}  \tau_{\mu}$ ($\tau_0  =$ unit matrix)  is a  bijection of
$\br_M^4$ and $H(2,\bc )$ such that $x^2 = \det (x_{\mu}\tau_{\mu} )$.
But  $\{  \tau_{\mu}\}$  are  matrices  whose  elements  are  gaussian
integers,  i.e., complex numbers  whose real  and imaginary  parts are
integers.  Therefore the restriction of $\br_M^4$ to $\bz_M^4$ gives a
bijective map of $\bz_M^4$ into $H (2,\bz [i])$ where $\bz [i]$ stands
for the ring of gaussian  integers.  The group $SL (2,\bz [i]) \subset
SL(2,\bc  )$ has an action on  $H  (2,\bz[i])$ by  $X_{\mu}\tau_{\mu}  \lr  A
(X_{\mu}\tau_{\mu})  A^*$,  preserving  the  determinant,  exactly  as
$SL(2,\bc )$ acts on $H(2,\bc  )$.  Hence there is a discrete Lorentz
transformation $\Lambda$  such that  $(\Lambda X)_{\mu} \tau_{\mu}  = A
(X_{\mu}\tau_{\mu} )A^*, ~~ (\Lambda X)^2  = X^2,$ with
$A \lr \Lambda$ defining
a  homomorphism  of  $SL  (2,\bz  [i])$  into  $SO(3,1,\bz  )  \subset
SO(3,1,\br )$.   But since $A \in  SL (2,\bc )$, a  continuity argument shows
[17] that $\Lambda$, as an element of $SO(3,1,\br )$, belongs to its
connected component and hence to $SO(3,1,\bz )/\{ \pm 1 \in SO(3,1,\bz )\}
= L(\bz )$.  It is easy to check now that the kernel of the homomorphism
$A \lr \Lambda$ is the centre of $SL (2,\bz [i]) = \bz_2 = \{ \pm 1 \in
SL (2,\bz [i] )\}$ and that $L(\bz )$ 
is the quotient of $SL(2,\bz [i])$ by its 
centre, but not a subgroup.  All this is exactly as for $L(\br )$ and 
$SL(2,\bc )$.  (In the rest of this paper, $\Lambda$ will denote both 
an element of $L(\bz )$ and the corresponding element(s) of $SL(2,\bz [i])$;
no confusion will arise).

$SL(2,\bz [i])$ is the sought for group $\wh{L}(\bz )$.
A representation of $\wh{L}(\bz )$ restricting to its centre $\bz_2$
as the trivial (nontrivial) character will pass to the quotient 
group $L(\bz )$ as a trivial (nontrivial) projective representation.

\newpage
\noindent
{\bf 3. Unitary Representations of $\wh{P}(\br )$ - an Overview}

For the eventual construction of unitary representations of $\wh{P} (\bz )$,
we shall try to follow Wigner's method for $\wh{P} (\br )$ [10-12,18]
in a variant form described in [19], somewhat further streamlined.  
The generality of the method [11,20] allows room for hoping that,
with suitable adjustments, it can be adapted to the discrete case.
More importantly, the method naturally highlights the physical 
attributes, mass and (Lorentz) helicity, that permit a direct association
of elementary quantum fields with irreducible unitary representations.
The brief recapitulation below of this method of ``inducing from 
little groups''
is meant to highlight criteria for picking out certain unitary 
representations of $\wh{P} (\br )$ as physical (or physically acceptable)
especially in the case of massless representations.  
It will serve as a model for deciding which representations of 
$\wh{P} (\bz )$ can be considered physical.

The momentum space is the dual group of the translation group $\br^4$,
isomorphic to $\br^4$.  We denote it by $M$.  Let $\co$ be an orbit of
$\wh{L} (\br )$ in $M$ (the action is defined to be the natural action 
of $L(\br )$, lifted to $\wh{L} (\br )$ by letting the central subgroup
$\bz_2$ act trivially) and $S$ the stabiliser (little group) of any
point in $\co: \co = \wh{L} (\br )/S$.  
We suppose given, to begin with, a {\it finite
dimensional} representation $\rho$ of $\wh{L}(\br )$ on a Hilbert space 
$V$ with the property that the restriction of $\rho$ to $S$ is unitary. 
Denote by $\pi$ the projection of 
$\wh{L} (\br )$ onto  $\co$ and let $\sigma$ be
a section of $\pi$, i.e. any map $\co \lr \wh{L} (\br )$ such that
$\pi (\sigma (p)) = p$ for all $p \in \co$ and $\omega$
the $\wh{L} (\br )$-invariant measure on $\co$.

On the space $\ch_{\co ,V}$ of vector-valued functions $\phi$,
$\psi : \co \lr V$, square-integrable with respect to $\omega$, 
define a bracket $\lan \phi ,\psi \ran$ by
$$ \lan \phi , \psi \ran = \int_0 d\omega (p) \lan \rho (\sigma (p)^{-1})
\phi (p), \rho (\sigma (p)^{-1})\psi (p)  \ran_{V}.$$
If $\sigma$ and $\sigma '$ are two sections of $\pi$, it follows from $\pi
(\sigma (p)) = \pi (\sigma '(p ))$ $(=p)$ that $\sigma (p)^{-1}
\sigma '(p)$ is in  $S$.  And since $\rho$ restricts to
$S$ as a unitary representation, $\lan ~,~ \ran$ is independent of the
section used to define it making it a scalar product.  Moreover, 
$|| \phi || =0$ if and only if $\rho (\sigma (p )^{-1})\phi (p) =0$
for all $p \in \co$  i.e., $\phi =0$ identically.  So (the 
completion of) $\ch_{\co , V}$
is a Hilbert space.  Noting that $p \lr \rho (\sigma (p)^{-1}) \phi (p)$
is a section of a vector bundle over $\co$ with fibre $V$, 
we can characterise $\ch_{\co ,V}$ as the Hilbert space of such sections
which are $L^2$ with respect to $\omega$.

On $\ch_{\co ,V}$, $\wh{P} (\br )$ has a unitary representation given by
$$(U_{\co ,V} (\lambda ,a )\phi )(p) = \chi_p (a) \rho (\lambda ) \phi
(\lambda^{-1} p), ~~\lambda \in \wh{L} (\br ), ~~ a \in \br^4,$$ 
where
$\chi_p (a) = \exp (ipa)$ is the character of $\br ^4$ corresponding 
to $p$.  We have
$$ || U_{\co ,V} (\lambda ,a) \phi ||^2 = \int d\omega (p)
|| \rho (\sigma (\lambda p )^{-1}) \rho (\lambda ) \phi (p) ||^2_V$$
using the invariance of the measure.  But $\sigma (\lambda p)$ and
$\lambda \sigma (p)$ have the same projection onto $\co$ and hence 
differ by an element of $S$, $(\sigma (p)^{-1}\lambda^{-1} \sigma (\lambda p)$
is a cocycle $\wh{L} (\br )\times \co \lr S$) implying
$$\rho (\sigma (\lambda p)^{-1}) \rho (\lambda ) = \rho (s)^{-1}
\rho (\sigma (p)^{-1})$$
for some $s \in S$.  Since $\rho$ is unitary on $S$ by assumption,
the unitarity of $U_{\co ,V}$ follows.  We shall say that $U_{\co ,V}$
is supported on $\co$ and ranges over $V$.  It 
is irreducible whenever $V$ is an irreducible 
representation of $\wh{L}(\br )$.  In the language of induced
representations, $U_{\co ,V}$ is the representation induced by
the (unitary) restriction of $\rho$ to $S \subset \wh{L} (\br )$.

When $\co$ is a positive (mass)$^2$ positive (or negative) energy
mass shell of mass $m$, 
i.e., the orbit $\co_m$ through $p = (m,0,0,0)$, the stabiliser $S_{m,p}$ 
of $p\in \co_m$ is isomorphic to $SU(2)$ and every irreducible representation
$\rho$ of $\wh{L} (\br )$ restricts to $S_{m,p}$ as an irreducible unitary
representation.  Consequently, the helicity spectrum of $U_{m ,V}$
is determined equivalently and alternatively by the $\wh{L} (\br )$
or $S_{m,p}$ transformation properties of the function $\phi$; in particular,
the number of helicity states in $\ch_{m ,V}$ is $\dim V$.  Moreover, the
condition that $S_{m,p}$ fixes $p$ translates as the condition
$$(U_{m,V} (s,a) \phi )(p) = \chi_p (a) \rho (s) \phi (p),$$
for all $s \in S_{m,p}$, on the functions $\phi$.  This, or rather its
Lie algebra version, is the invariant wave equation or the free field
equation corresponding to the unitary representation $U_{m,V}$ of 
$\wh{P} (\br )$ [21].

In the light of the fact that all elementary particles have (mass)$^2
\geq 0$ and a finite set of helicities, we shall in general refer to unitary
representations of $\wh{P} (\br )$ with these properties (in particular
$\dim V < \infty$) as physical.  The condition on the helicity spectrum is a
powerful one - whenever $S$ is a noncompact group, it puts strong
restrictions on the unitary representations of $S$ that can be used in the
induction procedure.  It will play a crucial role in sorting out
physical representations of $\wh{P} (\bz )$ as indeed it already
does for massless representations of $\wh{P} (\br )$.

There are three mass $=0$ orbits: the vertex of the light cone in $M$ (a
one-point orbit) and the open upper and lower half light cones.  Consider
representations supported on the upper half light cone $C_+$.  The stabiliser
$S_0$ is isomorphic to the subgroup of upper triangular matrices in
$SL (2,\bc )$ (the representative point of $C$ which $S_0$ fixes is
$(p_0,0,0,p_0)$ for any $p_0 > 0$) which we choose to parametrise as

$$s(\theta ,z ) = \pmatrix{\exp (i\theta ) & z \exp (-i\theta ) \cr
0 & \exp (-i\theta )}, ~~ 0 \leq \theta < 2\pi , ~~ z \in \bc ,$$

\noindent
so that $s(\theta_1,z_1) s(\theta_2,z_2) = s(\theta_1 +
\theta_2 (\mod ~2 \pi ), z_1 + z_2 \exp (2i\theta_1 ))$.
So $S_0$ is the euclidean group of the plane $E (2, \br ) = SO (2,\br )
\vec{\times}$ $\br^2$, with $\br^2 = \{ ({\rm Re}~z , {\rm Im}~z )\}$
on which $SO(2,\br )$ acts as the two-fold cover of the rotation group.

Now, a finite dimensional unitary representation of $E (2,\br )$ is
necessarily nonfaithful; in fact the only such representations are
characters of the subgroup $SO(2,\br )$ and have the normal subgroup
$\br^2$ as kernel [22].  Hence $\wh{L} (\br )$, being simple, cannot have
any finite dimensional representation restricting to $E (2,\br )$
unitarily.  This means that the straightforward induction procedure that
works for massive representations is no longer valid and has to be
modified suitably.  The well-known way to do this [21,12] is to 
replace the Hilbert space $\ch_{\co ,V}$ of functions 
$\phi : C_+ \lr V$ by a subspace $\ch'_{\co,V}$ on which $U$
restricted to the subgroup $\br^2$ of $E(2,\br )$ acts trivially:
$$\rho (r) \phi (r^{-1}p) = \phi (p),  ~r = ({\rm Re}~z,{\rm Im}~z) \in
\br^2.$$  
The Lie algebra form of this condition constitutes the 
subsidiary condition.  A character of $SO(2,\br ) = E (2,\br )/\br^2$,
$\rho (\theta ) \phi (p) = \exp {(im~\theta /2}) \phi (p)$,
$\theta /2  \in SO(2,\br )$, $m \in \bz$, then induces a unitary 
representation of $\wh{P} (\br )$ on $\ch_{\co ,V}'$.  The fact
to be emphasised, and relevant in the context of $\wh{P} (\bz )$, is
that massless finite helicity representations of $\wh{P} (\br )$
exist because $E (2,\br )$ has a normal subgroup with compact quotient
group.  A massless  physical
irreducible unitary representation of $\wh{P} (\br )$ has precisely
one (Lorentz) helicity,  namely the character of $SO(2,\br  )$ to which
$\rho$  restricts.  This  helicity  is not  related  to rotational  spin
which, in any case, is a meaningless notion for a state that cannot be
transformed to rest.

For the sake of completeness, it should be remarked that all (mass)$^2
<0$  representations  are  doubly  unphysical.   In  addition  to  the
well-known causality  problem, they suffer  from unphysical helicities
as  well: the  stabiliser, which  is $SL(2,\br  )$, has  no nontrivial
finite dimensional unitary representations at all.

\noindent
{\bf 4. Physical Masses and Helicities for $\wh{P} (\bz )$}

This section is  devoted to an examination of the  extent to which the
fundamental
notions of mass and helicity can  be carried over from $\wh{P} (\br )$
to $\wh{P}  (\bz )$, as a  prerequisite to the  construction of unitary
representations  of $\wh{P}  (\bz )$ which are physically acceptable.

The momentum space of discrete space time is the dual group of the
discrete  translation  group  $\bz^4$,  namely  the  4-torus  $\bt^4$,
denoted simply by $T$ from now  on.  It is convenient for what follows
to think of  $T$ as $M/\bz^4$, where $M$  is the momentum space
$\br^4$ of the  continuum translation group and $\bz^4$  is the reciprocal
lattice.   Introducing coordinates $\{p_{\mu}\}$  in $M$,  we identify
$T$  as the  hypercube $\{  -\pi \leq  P_{\mu} \leq  \pi$, ~  $P_{\mu} =
p_{\mu} ~ (\mod~2\pi)\}$,  i.e. as  the  fundamental  region for  the
translation  action  of $\bz^4$  on  $\br^4$,  the  unit cell  of  the
reciprocal lattice.  (Physically, of course, all this just means that 
momentum is defined and conserved modulo $2\pi$).
We can then study the action of $\wh{L} (\bz )$
on $T$ by starting with its action on $M$ and translating the image (of a 
point  in $T \subset M$) back  to $T$ by some  integral multiples of
$2\pi$.  Under  this projection $\tau$: $M  \lr T$, $P_{\mu}  = \pi$ and  
$P_{\mu} = -\pi$ get identified for every $\mu$.

\begin{figure}[htpb] 
\vskip 10truecm
{\includegraphics{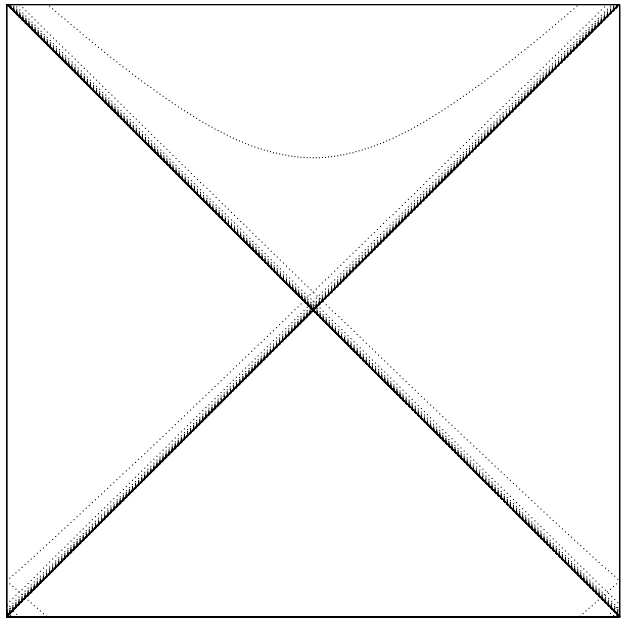}}
\caption{A typical ``torus mass-shell''.} 

\label{fig:ppd} 
\end{figure}

Thus every orbit  $\co_T$ of $\wh{L}(\bz )$  on $T$, through  a given point
$P$, can be determined by  first finding the orbit $\co$ of $\wh{L}
(\bz )  \subset \wh{L}(\br )$ in  $M$ through $P$  and then projecting
$\co$ back to $T$.  It is to be expected that, generically, such orbits
will be quite wild [23].  $\co$ being a subset of an orbit of
$\wh{L} (\br )$ in $M$, let us first determine the projection onto $T$
of  a positive mass  orbit $\co_{T,m}$ of $\wh{L}(\br  )$ 
in  $M$, the  familiar mass shell.  Figure ~\ref{fig:ppd} shows  such
an  orbit of  $\wh{L}  (\br  )$ in  $T$, projected further onto, say,
the (0,1) plane.  The corresponding orbit of $\wh{L} (\bz )$ for a
``mass'' $< \pi$ is a subset of this.

The  pathological nature  of positive  ``mass'' orbits  $\co_{T,m}$ is
made dramatically obvious by Figure~\ref{fig:ppd}.  Despite their being
the projections onto the unit  cell of physical positive mass orbits
of $\wh{L} (\bz )$ in $M$,  no fixed invariant mass  can be associated
to  them. (This is just a  reflection of the fact  there are no
$\wh{L} (\bz )$-invariant (and hence $\wh{L} (\br  )$-invariant)
nontrivial periodic functions on $M$ and stems from the periodicity of
the momentum itself).  On the contrary, $\co_{T,m}$ has points
corresponding to arbitrarily small positive ``(mass)$^2$'' as well as
tachyonic points with negative ``(mass)$^2$''   which  are   reached
by   large  boosts.  Representations supported  on such orbits will
violate (micro) causality and must be rejected.

On the  other hand,  consider the  upper half light  cone $C_+$  in $M$.
Under projection onto  $T$ (translation by multiples of  $2 \pi$), its
image $C_T$ is the whole of the light cone lying in $T$, including the
origin  and  the  negative energy  light  cone  inside  $T$.  $C_T$  is  a
hypersurface in $T$ which we shall refer to as the light cone of $T$.  We
can  consistently associate  a vanishing mass  to  every point  of $C_T$ 
--  the
polynomial $p_0^2-p_1^2-p_2^2-p_3^2$ is periodic and invariant as long
as it  vanishes.  The  orbit through any  point in $C_T$  of $\wh{L}
(\bz )$ is  a set of discrete points in  $C_T$ and every 
such orbit  can be said,
invariantly, to have zero mass.

Thus while  no physically sensible meaning  can be given  to a non-zero
mass, masslessness  is a  notion which makes  sense on the  lattice. 
Another way of understanding this distinction is to note that a massive
state  can be transformed  to rest  and then  subjected to  rotations in
order to determine its helicity.   But the lattice rotation group is a
finite group and cannot possibly serve to define arbitrary helicities.  A
massless state is free from this paradox.

The last statement  leads us naturally to the  problem of defining 
helicities in terms of  the lattice group $\wh{L}  (\bz )$. 
The situation here is as nice as it can be. We have the key result:
\begin{center}
\begin{tabular}{l}
  {\it Every finite  dimensional irreducible representation of $L(\br)$
or $\wh{L}
    (\br )$  restricts to its lattice} \\
{\it subgroup $L(\bz )$ or
$\wh{L} (\bz  )$ respectively as an
    irreducible representation.}
\end{tabular}
\end{center}

This is a special case of a general theorem, the density theorem of A.
Borel  [24], on  representations of  discrete subgroups  of noncompact
semisimple  Lie groups  [23-28].  
A general formulation of the theorem is the following.  Let $G$ be a
semisimple Lie group none of whose factors is compact and $\ct$ a discrete 
subgroup of $G$ having the property that $G/\ct$ has finite volume.  
Then every finite dimensional irreducible representation of $G$ remains
irreducible as a representation of $\ct$.  The discrete groups 
$SO(3,1,\bz )$ and $SL (2,\bz [i])$ have finite covolumes in 
$SO(3,1,\br )$ and $SL(2,\bc )$ respectively and hence meet the
conditions of the theorem.  

Thus the helicity content of an irreducible finite dimensional
representation of $\wh{L} (\br )$ remains intact on restricting 
$\wh{L} (\br)$ to $\wh{L} (\bz )$; nothing is lost in this regard by 
discretising space-time as long as we do not try to define
helicities through the rotation group.
With  this  result in  hand, we  can
attempt the construction  of irreducible unitary representations of
$\wh{P} (\bz )$, characterised by a mass $=0$ and a helicity identical
to that corresponding to an irreducible representation of $\wh{L} (\br
)$.   Moreover, from the density  theorem, 
the relationship between finite dimensional
representations of $L(\bz  )$ and $\wh{L} (\bz )$  is exactly the same
so that between representations of $L (\br )$ and $\wh{L} (\br )$:
the linear representations of $\wh{L} (\bz )$ which are trivial on its 
centre and hence pass to linear representations of $L(\bz )$ are integral
helicity representations and those which are not, and hence pass to
nontrivial projective representations, are half-odd-integral 
helicity representations.
In particular, $\wh{L} (\bz )$ has ``spin $\frac{1}{2}$'' representations
of both chirality and chiral Weyl spinor fields on $\bz_M^4$ 
can naturally be associated with them.

The validity of the density theorem for the spin $\frac{1}{2}$ 
representations, the most important in practice, is actually 
easy to establish.  The defining (left-chiral) representation
$\rho_L$ of $SL (2,\bc ) : \rho_L (\lambda ) = \lambda \in
SL (2,\bc )$, restricts to $SL (2,\bz [i])$ as $\rho_L (\Lambda ) =
\Lambda \in SL (2,\bz [i])$.  Let $B$ be any operator on $\bc^2$, the 
representation space of $\rho_L$, commuting with $\rho_L (\Lambda )$ for all
$\Lambda \in SL (2, \bz [i])$.  The Pauli matrices $\{ \tau_i\}$ multiplied
by $i$ are obviously in $SL (2, \bz [i])$.  Hence $B$ commutes with 
$\rho_L (i \tau_i ) = i \tau_i$ by assumption, which means that $B$ is a
multiple of the unit operator.  By Schur's lemma, $\rho_L$ restricted
to $SL (2,\bz [i])$ is thus irreducible.  The argument for the conjugate
right-chiral representation is the same.  (We may note that any two of
the Pauli matrices can be chosen to belong to a set of generators for
$SL (2,\bz [i])$ [29]).

We  shall  consider  as  physical  those and  only  those  irreducible
representations  of   $\wh{L}  (\bz   )$  that  are   restrictions  of
(continuous) finite dimensional irreducible representations of $\wh{L}
(\br )$.

\newpage
\noindent
{\bf 5. Unitary Representations of $\wh{P}(\bz )$ - First Steps}
  
  The  continuum Lorentz group  $\wh{L} (\br  )$ acts  transitively on
  $C_+$, the  upper half light  cone in $M$  (and likewise on  $C_-$). 
  Thus all of $C_+$ is  one orbit of $\wh{L}$ and, consequently, there
  is an irreducible unitary representation of $\wh{P} (\br )$ supported
  on  $C_+$ and  ranging over  a given  irreducible  representation of
  $\wh{L} (\br )$ (subject, of  course, to the subsidiary conditions). 
  This situation fails to hold for the action of $\wh{L} (\bz )$ on the
  torus  light  cone  $C_T$,  there  being a  continuous  infinity  of
  disjoint orbits.  A theory of unitary representations of $\wh{P}(\bz
  )$  \`a la  Mackey,  physically acceptable  in  the sense  explained
  earlier,  would   require  a  characterisation  of   the  orbits  of
  $\wh{L}(\bz )$  in $C_T$, the determination of  their stabilisers and,
  finally,  a further  classification  according to  the existence  or
  otherwise  of  finite  dimensional  unitary representations  of  the
  latter.  Needless  to say,  that will  be a challenge  and it  is not
  undertaken here.   We shall  content ourselves in  this preliminary
  look at the problem by exhibiting certain special types of orbits, of
  which one  type does not admit physical  representations and another
  type does, giving representations which are the discrete analogues of
  the  physical  massless representations  of  $\wh{P}  (\br )$.   

The origin in $C_T$  is a one-point orbit whose  stabiliser is the whole
  of $\wh{L} (\bz )$; it need not be considered further.

In general, we can find the orbit of $\wh{L} (\bz ) $ in $C_T$ through 
a point $P \neq 0$ by first finding the orbit in $C_+$ (or $C_-$)
through $P$ considered as point in $C_+$ and then mapping it to $C_T$
{\it via} the projection $\tau : C_+ \lr C_{T} = C_+/\bz^4$ (see
section 4).  We call $P$ a rational point if its coordinates are all
rational multiples of $\pi : P_{\mu} = (q_{\mu}/d_{\mu}) \pi$, $q_{\mu}$,
$d_{\mu} \in \bz$ with $-|d_{\mu}| \leq q_{\mu} \leq |d_{\mu}|$ for
$\mu = 0,1,2,3,$ and an irrational point otherwise.  Every point in the 
orbit through a rational (irrational) point is rational (irrational).

Consider  rational  orbits   first.   Reexpress  the  coordinates  $\{
q_{\mu}/d_{\mu}\}$ (dropping  the factor $\pi$  for the time  being) in
terms of  the lowest positive  common multiple $D$ of  $\{ d_{\mu}\}$,
i.e., $P_{\mu} = Q_{\mu}/D$ with $-D \leq Q_{\mu} \leq D$, and no positive
integer  $D'  <  D$ exists  such  that  $P_{\mu}  = Q_{\mu}'/D'$  for  any
$\{Q_{\mu}'\}$ with  $-D' \leq  Q_{\mu}' \leq D'$  for all  $\mu$; $\{
Q_{\mu}/D\}$ will be referred as the standard expression for $P$.  Then
$\Lambda  \in \wh{L}  (\bz )$  acts on  $P$ by  changing  its standard
expression to $\{ \Lambda_{\mu \nu}Q_{\nu} /D\}$.  Suppose $\Lambda_{\mu \nu}
Q_{\nu}$  has a  common  factor with  $D$  for each  $\mu$.  Then  the
standard   expression  for   $\Lambda  P$   will  have   a  denominator
$D_{\Lambda}$  strictly less  than $D$;  otherwise $D_{\Lambda}  = D$.
But  the  same  argument  applies  also to  $\Lambda^{-1}$  acting  on
$\Lambda P$,  implying that $D  \leq D_{\Lambda}$.  So  $D_{\Lambda} =
D$; $\wh{L} (\bz  )$ acts on every rational  point without changing the
denominator  in   its  standard  expression.    Since  the  numerators
$Q_{\mu}$ are  bounded between $-D$  and $D$,  we conclude  that every
rational orbit of $\wh{L} (\bz )$ in $C_T$ is a finite set.
 
The stabiliser  in $\wh{L} (\bz )$  of any rational point  of $C_T$ is
therefore a  subgroup of  finite index, in  other words almost  all of
$\wh{L} (\bz  )$.  The situation  is practically identical to  that of
the one-point orbit  consisting of the origin -  the stabiliser has no
finite  dimensional  unitary   representation  from  which  a  unitary
representation  of $\wh{P}  (\bz )$  with a  finite  helicity spectrum
(subject to a finite set of subsidiary conditions, see remarks in section
3 on massless representations of $\wh{P} (\br ))$ can be induced.  So 
rational orbits are to be discarded.

We turn now to irrational orbits.  Consider in particular the orbit 
through a point with momentum along an axis, say $P = (P_0, P_0, 0,0)$,
with $P_0$ irrational.  For the action of $\wh{L} (\bz )$ on $T$, 
its stabiliser is 
$$ \Sigma_P = \{ \Lambda \in \wh{L} (\bz )~|~ (\Lambda P)_{\mu} = P_{\mu}
~~ {\rm mod}~ \bz \},$$
i.e., the subgroup satisfying the conditions 
$$\begin{array}{rcl}
(\Lambda_{00} - \Lambda_{01}) P_0 & = & P_0 + N_0, \\
(\Lambda_{10} - \Lambda_{11}) P_0 & = & P_0 + N_1, \\
(\Lambda_{20} + \Lambda_{21}) P_0 & = & N_2, \\
(\Lambda_{30} + \Lambda_{31}) P_0 & = & N_3,
\end{array}$$
for arbitrary integers $N_0, N_1, N_2, N_3$.  These conditions 
can hold for irrational $P_0$ only if all the $N$ vanish. It
follows that the stabiliser of $P = (P_0,P_0,0,0)$, $P_0$ irrational,
for the $\wh{L} (\bz )$ action on $C_T$ coincides with its stabiliser
for the $\wh{L} (\bz )$ action on the whole of $C_+$.  The latter
can be found exactly as in the case of $\wh{L} (\br )$, leading to
the conclusion that $\sum_P$ is the subgroup of $SL (2,\bz [i])$
consisting of upper triangular matrices
$$s (\zeta ,Z ) = \pmatrix{\zeta & \zeta^{-1} Z\cr
0 & \zeta^{-1}}, ~ \zeta ,\zeta^{-1}, Z \in \bz [i].$$
The only elements of $\bz [i]$ with inverses in $\bz [i]$,
namely its units, being $\zeta = \pm 1, \pm i$, the subgroup
of diagonal matrices
$$s (\zeta , 0) = \pmatrix{\zeta &  0 \cr 0 & \zeta^{-1}}$$
is the cyclic group $\bz_4$.  The subgroup of elements $s (1,Z )$
is the planar lattice $\bz^2$ whose points are identified with
$({\rm Re}~Z , {\rm Im}~Z)$.  The multiplication in $\Sigma_P$ is
given by
$$ s(\zeta_1 , Z_1) s (\zeta_2  ,Z_2) = s(\zeta_1 \zeta_2, Z_1+\zeta_1^2 Z_2)$$
confirming that $\Sigma_P$ is indeed the discrete euclidean group
$E(2,\bz ) = \bz_4 \vec{\times} \bz^2 = SO(2,\bz ) \vec{\times} \bz^2$,
with $\zeta \in \bz_4$ acting on $\bz^2$ by $({\rm Re}~Z ,{\rm Im}~Z)
\lr ({\rm Re}~\zeta^2Z,{\rm Im}~\zeta^2 Z$) - the action by $\zeta^2$ is
a reminder that $\wh{L} (\bz )$ covers $L(\bz )$ twice, again as in the
continuum theory.

The reasoning given above covers all orbits containing a point of the
form $P = (P_0,P_0,0,0)$, $(P_0,0,P_0,0)$ or $(P_0, 0,0, P_0)$ with $P_0$
an arbitrary irrational number between $-1$ and 1.  Given such an orbit
$\co$, which is a discrete set, we can define (square summable) functions 
on it with values in a finite dimensional irreducible representation space 
$V$ of $\wh{L} (\br )$ (and hence, by the density theorem, of $\wh{L} (\bz )$)
and carry through the Wigner-Mackey construction of physically acceptable 
unitary representations of $\wh{P} (\bz )$.  This is made possible precisely 
because the stabiliser $E(2,\bz )$ has finite dimensional unitary 
representations, those that are trivial on its $\bz_2$ subgroup.
However, it is not true in general that light-like momenta along different
spatial directions, or along the same direction but are linearly independent
over rationals, can be connected by discrete Lorentz transformations.

The actual construction of  the unitary representation of $\wh{P} (\bz
)$ supported on  an irrational orbit $\co$ having  $\Sigma = E(2,\bz)$ as
stabiliser  and  ranging  over  $V$  may, finally,  be  summarised  as
follows.  Define  the momentum space  fields as functions $\phi  : \co
\lr V$ satisfying the subsidiary condition
$$\rho (R) \phi (R^{-1}P) = \phi (P)$$
where  $\rho$ is  the irreducible
representation of  $\wh{L} (\br  )$ (and  hence, to
repeat, of  $\wh{L} (\bz )$) on $V$  and $R$ is any  element of $\bz^2
\subset E(2,\bz )$.  If $\sigma$ is a section $\co \lr \wh{L} (\bz )$,
such fields  form a Hilbert space $\ch_{\co ,V}'$
with scalar product
$$\langle \phi ,\psi \ran = \sum_{p \in \co} \lan \rho (\sigma (P)^{-1})
\phi (P), \rho (\sigma (P)^{-1}) \psi (P) \rangle_V$$
if $\rho$ restricted to $E (2,\bz )$ is a unitary character of $SO(2,\bz )$,
(assuming of course that $\langle \phi | \phi \rangle < \infty$).  On 
$\ch_{\co ,V}'$ we have a unitary representation of $\wh{P} (\bz )$ given
by the action
$$(U_{\co,V} (\Lambda ,A) \phi )(P) = \exp (iP_{\mu}A_{\mu}) \phi 
(\Lambda^{-1} P),~~\Lambda \in \wh{L} (\bz ), ~ A \in \bz^4.$$

All  this is a  direct adaptation  of the  standard method  familiar from
continuum  relativity, with  the notable  difference arising  from the
proliferation of even  good (irrational) orbits of $\wh{L}  (\bz )$ in
$C_T$.   The  main   remaining  task  is  thus  to   put  the  unitary
representations supported on each of these good orbits together, so as
to obtain  one which is  supported on all  of $C_T$ and  is physically
acceptable.  Once that is done,  the rest of the canonical theoretical
framework --  Fourier-transforming the  momentum space fields  to the
space - time lattice, writing down  the field equations,  etc. can be
completed.

\noindent
{\bf 6. What Next?}

>From the  viewpoint of approximating continuum field  theories, the key
advantage  of  minkowskian  over  euclidean lattices  is  clearly  the
possibility of defining helicities  in a satisfactory way for massless
fields.  Chiral fields of any  helicity are the natural inhabitants of
minkowskian lattices,  in sharp  contrast to  the  situation for
euclidean lattices, where  a general theorem [30] says  in effect that
chiral  fermion fields  cannot  be defined;  indeed, chiral  symmetry
itself is not easy to define [31,32].  Since the crucial first step in
defining chiral  fields is to  have a good  notion of helicity,  it is
reasonable  to expect  that this  phenomenon reflects  the fundamental
difference in the discrete subgroups of  $\wh{L} (\br ) = SL (2,\bc )$
and the  corresponding euclidean group  Spin$(4,\br )$ ,  namely that
$SL (2,\bz [i])]$ is a  ``dense'' subgroup of $SL(2,\bc )$ while Spin
$(4,\bz  )$ is  the finite  group $SU (2 ,\bz  [i]) \times  SU (2,\bz
[i])$.  $(SU  (2,\bz [i])$  is the 8th  order group whose  elements are
$\pm  1,  \pm  i  \tau_i$  $(\{ \tau_i\}  =$  Pauli  matrices)).   This
expectation turns out to be justified (as will be shown separately).
It is also to be kept in mind that in the usual treatment [33] lattice 
fermion fields $\psi (X)$ are taken to satisfy (Dirac or Weyl) equations
of motion corresponding to a $\psi$ transforming by the continuum 
Lorentz group while its argument can only transform by the lattice
group.  The consequences of this inconsistent procedure become manifest
in any attempt at its physical interpretation [34].

Next, still on the  positive side, it is entirely  straightforward to define
gauge fields,  abelian or nonabelian, on a  minkowskian lattice, exactly
as in  the euclidean case [4],  and to make models  of massless chiral
fermions interacting with them, e.g., a lattice version of the unbroken
standard  model.  On the  negative side,  however, breaking  gauge and
chiral  symmetry on the  lattice, so  as to  generate gauge  boson and
fermion masses  would appear to be  a nontrivial task, in  view of the
exact  masslessness of  physically acceptable  representations of $\wh{P}
(\bz )$.  One possibility is that the generation of vacuum expectation
values of scalar  fields is a local phenomenon,  i.e., that the vacuum
becomes degenerate,  or that some  order parameter becomes nonzero, in the
continuum limit.  Alternatively, it may  be that the breaking of gauge
symmetry on  the lattice is accompanied by  the simultaneous breaking
of strict  $\wh{P} (\bz )$ invariance  so as to  allow nonzero masses
to emerge.

Apart form gauge model building,  and even more speculatively, one can
wonder whether minkowskian lattices can  serve as a general model for
space-time at a sufficiently small  length scale (the Planck length?). 
Natural  deviations from special  relativistic symmetries  will become
operative at that scale, on account of gravitational effects - one cannot 
then talk meaningfully of discrete Poincar\'e invariance in isolation.  
Nonzero 
masses are, perhaps, a general-relativistic artifact and the ``real world''
a coarse-grained version of an underlying discrete space-time, the 
coarse-graining  being accomplished by gravitational interactions.

\vspace{4mm}
\noindent
{\bf Acknowledgements.} This work originated in discussions with 
M.S. Narasimhan and its development owes a great deal to the
frequent and continuing advice of M.S. Raghunathan on its 
mathematical aspects.  Warmest thanks are expressed to both of
them.  The interest and help of V. Chandrasekhar, N.D. Hari Dass,
Sourendu Gupta, Parameswaran Sankaran and T.N. Venkataramana are
gratefully acknowledged. 

\newpage
\noindent
{\bf References and Footnotes}

\begin{enumerate}
\item G. Wentzel, Helv. Phys. Acta {\bf 13}, 269 (1940), 
\item H.S. Snyder, Phys. Rev. {\bf 71}, 38 (1947).
\item L.I. Schiff, Phys. Rev. {\bf 92}, 766 (1953).
\item K.G. Wilson, Phys. Rev. D {\bf 10}, 2445 (1974).
\item R. Balian, J-M Drouffe and C. Itzykson, Phys. Rev. D {\bf 10},
3376 (1974).
\item D. Weingarten, Phys. Lett. {\bf B90}, 280 (1980).
\item D. Weingarten, Nucl. Phys. {\bf B210}, 229 (1982).
\item M. Creutz, L. Jacobs and C. Rebbi, Phys. Rev. Lett.
{\bf 42}, 1390 (1979).
\item Two sources (among others) of references to all but the 
most recent work are I. Montvay and G. M\"unster, {\it Quantum Fields on a 
Lattice} (Cambridge University Press, Cambridge, 1994) and J. Ambjorn,
B. Durhuus and T. Jonsson, {\it Quantum Geometry} (Cambridge University
Press, Cambridge, 1997).  Work directly relevant to the concerns of the
present paper will be individually cited at the appropriate place.
\item E. Wigner, Ann. Math. {\bf 40}, 149 (1939).
\item G.W. Mackey, Ann. Math. {\bf 55}, 101 (1952), {\bf 58}, 143 (1953).
\item V.S. Varadarajan, {\it The Geometry of Quantum Theory} (Springer, 1985).
Chapter XII contains an account of Wigner's construction from the viewpoint
of Mackey's general theory of unitary representations of semidirect product
groups as induced from appropriate little groups, including the derivation
of the field equations.
\item S. Weinberg, {\it The Quantum Theory of Fields, Vol I} (Cambridge
University Press, Cambridge, 1995).  Among the numerous texts on 
relativistic quantum field theory, this is the one that is most
faithful to Wigner's vision as expressed in [10].
\item M.S. Raghunathan, Rev. Math. Phys. {\bf 6}, 207 (1994) is a systematic
exposition of the theory of universal central extensions of (not
necessarily Lie) groups, perhaps the only such account directly 
accessible for use in quantum mechanics, see [15,16] below.  Instances of 
how the unjustified replacement of the universal extension of a 
Lie group by its 
universal cover can lead to fallacious physics 
will be found in these references.
\item P.P. Divakaran, Phys, Rev. Lett. {\bf 79}, 2159 (1997).
\item P.P. Divakaran, Rev. Math. Phys. {\bf 6}, 167 (1994).
\item R.F. Streater and A.S. Wightman, {\it PCT, Spin and Statistics 
and all that} (Benjamin, New York, 1964).
\item We may make what we can of the sentence ``$\cdots$ more general
equations, as far as they exist (e.g., in which the coordinate is quantized,
etc.) are also included in the present treatment'' from the introductory 
section of [10].
\item V.M. Niederer and L.O'Raifeartaigh, Fort. der Phys. {\bf 22}, 111
(1974).
\item G.W. Mackey, {\it Unitary Group Representation in Physics, 
Probability and Number Theory}, (Benjamin/Cummings, Reading, 1978).
\item V. Bargmann and E.P. Wigner, Proc. Nat. Acad. Sci. {\bf 34}, 211
(1948).
\item A thorough account of the unitary representation theory of $E (2,\br)$
is available in M. Sugiura, {\it Unitary Representations and Harmonic 
Analysis} (2nd ed) (North Holland/Kodansha, Amsterdam/Tokyo, 1990).
\item The study of the action of discrete subgroups of Lie groups,
especially semisimple Lie groups, on manifolds is a flourishing
subject and can potentially be of benefit to the physics of
discrete special relativity.  An accessible account can be found in R.J.
Zimmer, {\it Ergodic Theory and Semisimple Lie Groups} 
(Birkh\"auser, Boston, 1984).
\item A. Borel, Ann. Math. {\bf 72}, 179 (1960).
\item Work subsequent to [24] treats the subject with varying
degrees of generality and abstraction.  Of these, the treatments 
closest to our use of it are in [26,27].
\item M.S. Raghunathan, {\it Discrete Subgroups of Lie Groups}
(Springer, New York, 1972).
\item S.G. Dani, Math. Zeit {\bf 174}, 81 (1980).
\item For the sake of bibliographic completeness, it should be mentioned 
that the most comprehensive and general account of the subject is to
be found in G.A. Margulis, {\it Discrete Subgroups of Semisimple Lie Groups}
(Springer, Berlin, 1991).
\item R.G. Swan, Adv. Math. {\bf 6}, 1 (1971) lists a complete set of 
generators and a complete set of relations among them for $SL (2,\bz [i])$.
\item H.B. Nielsen and M. Ninomiya, Nucl. Phys. {\bf B185}, 20 (1981);
{\bf B193}, 173 (1981).
\item P.H. Ginsparg and K.G. Wilson, Phys. Rev. D {\bf 25}, 2641 (1982).
\item M. L\"uscher, Phys. Lett. {\bf B428}, 342 (1998).
\item The current status of the subject is reviewed in F. Niedermayer,
Nucl.  Phys. (Proc. Suppl.) {\bf 73}, 105 (1999).
\item For an exposition of the physics of the standard lattice fermion
theory, see N.D. Hari Dass, Int. J. Mod. Phys. {\bf B14}, 1989 (2000).
\end{enumerate}

\end{document}